\documentclass[]{eptcs}
\providecommand{\event}{QAPL 2014} 
\providecommand{\publicationstatus}{Extended abstract for a talk given at \event.}
\usepackage{breakurl}              
\usepackage{graphicx}
\usepackage{paralist}
\title{Quipper: Concrete Resource Estimation in Quantum Algorithms}
\author{
  Jonathan M. Smith
  \institute{University of Pennsylvania,\\Philadelphia, U.S.A.}
  \and
  Neil J. Ross\qquad Peter Selinger
  \institute{Dalhousie University,\\Halifax, Canada}
  \and
  \hspace{-1cm}Beno\^{i}t Valiron
  \institute{\hspace{-1cm}PPS, UMR 7126, Univ Paris Diderot,\\
    \hspace{-1cm}Sorbonne Paris Cit\'{e}, F-75205 Paris, France}}
\def\titlerunning{Concrete Resource Estimation in Quantum Algorithm}
\def\authorrunning{N.J. Ross, P. Selinger, J.M. Smith and B. Valiron}

\newcommand{\spacing}{-2ex}

\begin{document}
\maketitle

\begin{abstract}
  Despite the rich literature on quantum algorithms, there is a
  surprisingly small amount of coverage of their concrete logical
  design and implementation. Most resource estimation is done
  at the level of complexity analysis, but actual concrete numbers (of
  quantum gates, qubits, {\it etc.}) can differ by orders of
  magnitude. The line of work we present here is a formal framework to
  write, and reason about, quantum algorithms. Specifically, we designed a
  language, Quipper, with scalability in mind, and we are able to
  report actual resource counts for seven non-trivial algorithms
  found in the quantum computer science literature.
\end{abstract}

\vspace{\spacing}
\section{Introduction}
Quantitative analysis of programs implies a tool base with which
resource estimates can be made. In the case of a quantum programming language,
the relevant units are ultimately quantum bits (qubits), protected by 
complex error correction codes. The role of a quantum programming language
in the larger context of quantum computing is discussed by van Meter and
Horsman~\cite{Blueprint2013}. We implemented a quantum programming language,
Quipper~\cite{pldi,rc,quipper}, which is sufficiently complete that we can compile a quantum
algorithm from a high-level description into a low-level logical
quantum circuit, i.e., a sequence of quantum gates. From such a gate
sequence, the resources required by a quantum computation can be estimated.
We have performed resource estimates for seven quantum
algorithms, with problem sizes selected to be at the point where the
quantum algorithm should outperform a classical computer. The
power of our approach is that it carries with it
the generality of a programming language, needing only to be parameterized
by the characteristics of the problem. The resource estimates have also
identified ripe optimization targets for enhancements to Quipper.

\vspace{\spacing}
\section{Quantum computation}

Quantum computation deals with data encoded on the state of particles
governed by the laws of quantum physics. A piece of quantum data can be
seen as a complex combination (or {\em superposition}) of pieces of 
classical data. This is reminiscent of probabilistic computation, with 
the difference that the coefficients are complex numbers.

The most common model for a quantum computer is Knill's QRAM 
model~\cite{Knill-1996}, where a quantum device serves as co-processor 
to a classical unit.
\begin{center}
  \includegraphics[width=2.5in]{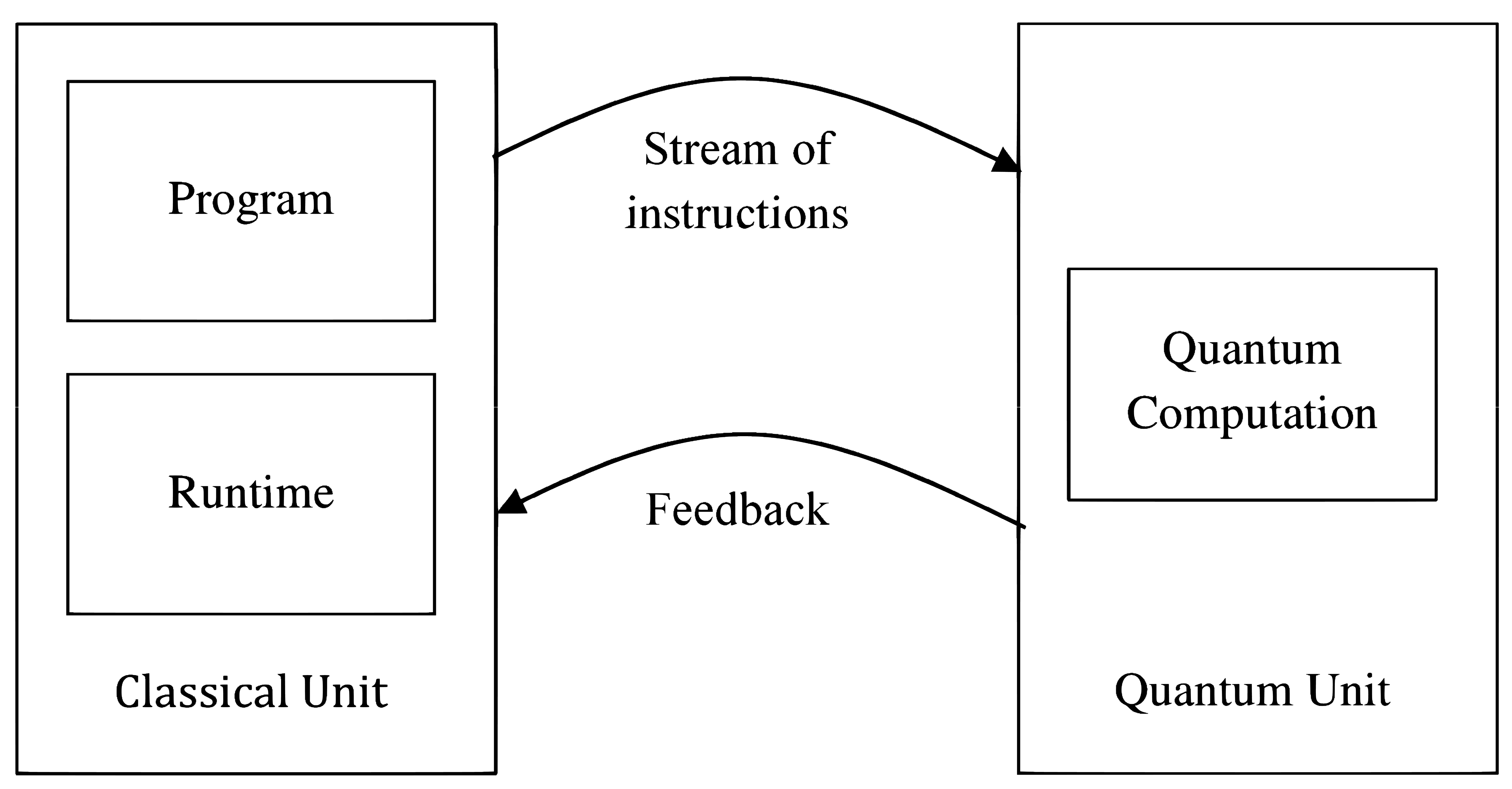}
\end{center}
The classical unit performs tasks such as compilation and bookkeeping, 
and can also send streams of instructions to the quantum unit, which 
only performs purely quantum operations. The instructions sent to the 
quantum co-processor are of three kinds:
\begin{compactitem}
\item initializations (to send classical data to the quantum device),
\item unitaries (which are reversible operations), and
\item measurements (to retrieve classical information from the quantum 
device).
\end{compactitem}
Measurements are the only way for the quantum device to provide 
feedback to the classical unit. This operation is probabilistic 
and can globally affect the state of the quantum device.

\vspace{\spacing}
\section{Generalized circuit model}

There is no control flow on the quantum co-processor. Any loop or 
conditional branching has to come from the classical device 
controlling the co-processor. As a result, a quantum computation can 
be pictured as a linear circuit, representing the flow of elementary 
instructions sent to the co-processor. In this representation, a wire 
stands for a quantum register (i.e., a {\em quantum bit}), and  a box 
represents an elementary operation.

Many algorithms use the quantum device in a simple, batch-style
fashion as follows:
\begin{center}
  \includegraphics[width=4in]{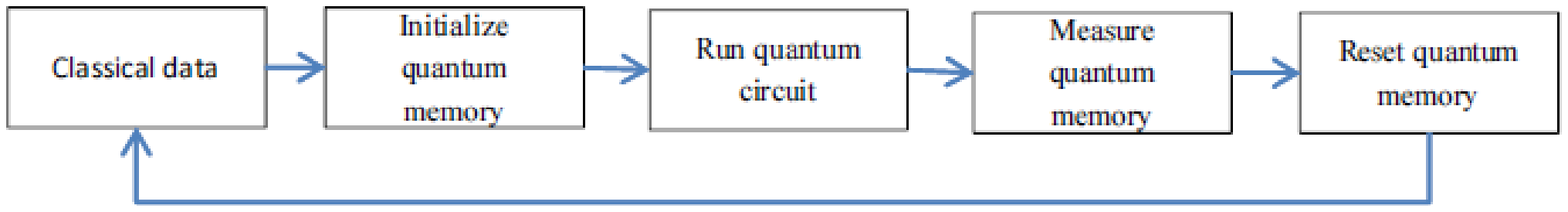}
\end{center}
In some algorithms, however, the circuit is conditioned on the result 
of intermediary measurements. Such a circuit is generated ``on the fly'' 
by the classical device, and a particular part of the circuit can depend 
on a measurement done at a previous stage:
\begin{center}
  \includegraphics[width=4in]{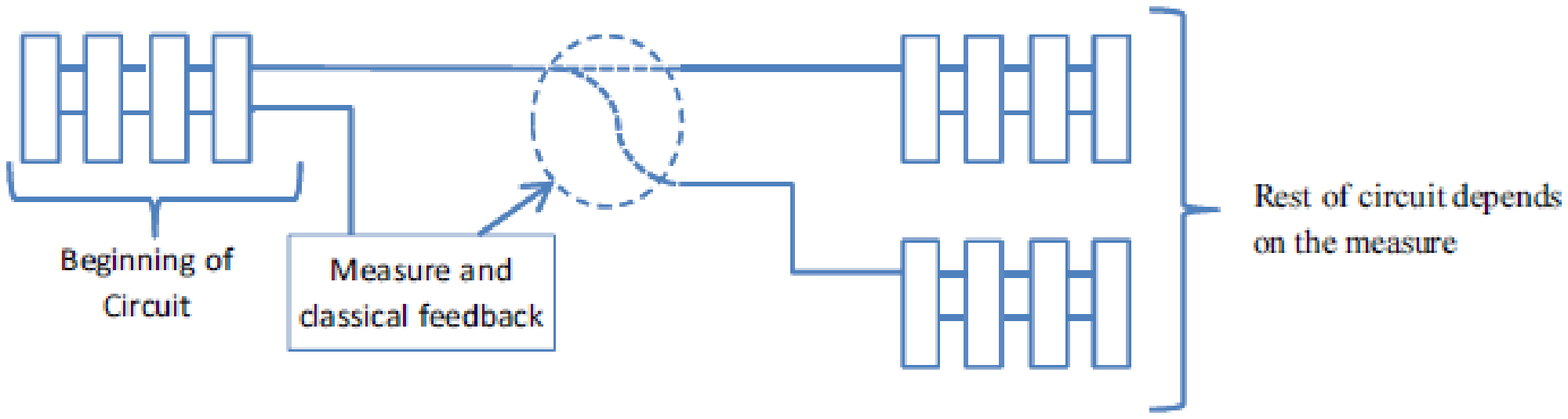}
\end{center}
A scalable quantum programming language must therefore accommodate such 
a dynamic representation of circuits.

\vspace{\spacing}
\section{Our proposal: Quipper}

We introduce Quipper~\cite{pldi,rc,quipper}, a functional language for quantum
computation embedded in Haskell. Quipper is intended to offer a
unified general-purpose programming framework for quantum
computation. Its main features are:
\begin{compactitem}
\item An extended circuit model. Initializations and terminations of
  qubits are tracked for the purpose of ancilla
  management.
\item Hierarchical circuits. Quipper features subroutines (or {\em
    boxing}) at the circuit level. This permits compact representation
  of circuits in memory.
\item A circuit description language. It can handle procedural and
  applicative paradigms of computation, and its monadic semantics
  allows high-level manipulations of circuits with programmable
  operators.
\item Two run-times. A Quipper program describes a
  {\em family} of circuits, which may depend on some classical
  parameters. After compilation, a program is first executed to
  generate a circuit (circuit generation time), and then the circuit
  is executed (circuit execution time).
\item Parameter/input distinction. Quipper has two notions of
  classical data: parameters, which are known at circuit generation
  time,  and inputs, which are known at circuit execution time. For
  example, the type {\tt 
    Bool} stands for parameters and the type {\tt Bit} for inputs.
\item Extensible datatypes. Quipper offers an abstract view of the
  notion of quantum data using the powerful type class mechanism of
  Haskell's type system.
\item Automatic generation of quantum oracles. Concrete quantum
  algorithms come with non-trivial classical operations that have to
  be lifted to quantum operations. Quipper comes with a facility to
  turn ordinary Haskell programs into reversible circuits, using
  Template Haskell.
\end{compactitem}
Similarly to what is done in Lava~\cite{lava}, the semantics of
circuit generation is captured in a monad. This permits both
imperative-style programming, where a circuit is described
gate by gate, and declarative-style, where circuits are manipulated
as first-class objects, and transformed with combinators.

An example of code with the corresponding circuit is shown below.

\medskip
\begin{tabular}{@{}ll@{}}
  \begin{minipage}{0.45\linewidth}
{\scriptsize\begin{verbatim}
circ :: Qubit -> Circ Qubit
circ x = do
    hadamard_at x
    with_ancilla $ \y -> do
       qnot_at y
       qnot x `controlled` y
       qnot_at y
    hadamard_at x
    return x
\end{verbatim}}
  \end{minipage}
  &
  \begin{minipage}{0.45\linewidth}
    \hspace{-3cm}\includegraphics[width=3in]{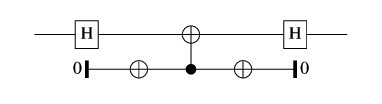}
  \end{minipage}
\end{tabular}

\vspace{\spacing}
\section{Achievements}

The language Quipper has been designed in the context of the
IARPA-funded QCS program {\cite{BAA}}.
Seven algorithms were used as benchmarks. These algorithms were
chosen by IARPA to
provide a reasonably representative cross-section of current
algorithms.
\begin{compactitem}
\item Binary Welded Tree (BWT). To find a labeled node in a
  graph~\cite{BWT}.
\item Boolean Formula (BF). To evaluate a NAND formula~\cite{BF}. The
  version of this algorithm implemented in Quipper computes a winning 
  strategy for the game of Hex.
\item Class Number (CL). To approximate the class group of a real
  quadratic number field~\cite{CN}.
\item Ground State Estimation (GSE). To compute the ground state
  energy of a particular molecule~\cite{GSE}.
\item Quantum Linear Systems (QLS). To solve a linear system of
  equations~\cite{LS}.
\item Unique Shortest Vector (USV). To choose the shortest vector
  among a given set~\cite{SV}.
\item Triangle Finding (TF). To exhibit a triangle inside a dense
  graph~\cite{TF}.
\end{compactitem}
These algorithms use of a wide variety of quantum primitives,
such as amplitude amplification, quantum walks, the quantum Fourier
transform, and quantum simulation. Several of them also require the 
implementation of complex classical oracles. The starting
point for each of our algorithm implementations was a detailed
description of the algorithm provided by IARPA. They were all coded in
Quipper and are running, in the sense that one can generate the
circuit (and portion thereof). 

Using Quipper, we were able to perform semi- or completely
automated logical gate count estimations for each of the algorithms. For example, in the case
of the triangle finding algorithm,
\begin{verbatim}
./tf -f gatecount -o orthodox -l 31 -n 15 -r 6
\end{verbatim}
produces the gate count for the complete algorithm. This runs to
completion in under two minutes on a laptop computer, and produces a 
count of 30,189,977,982,990 (over 30 trillion) total gates and 4676 
qubits. 

\vspace{\spacing}
\section{Conclusion and future work}

Quipper is a scalable language able to manipulate ``realistic'' 
quantum algorithms and input sizes. The process of compiling Quipper 
into a quantum circuit has provided fascinating quantitative insights 
into the hardware required to execute a quantum computation. As shown 
above with the triangle finding algorithm, the gate counts are large 
and may present a fundamental barrier to quantum computing unless 
significant optimizations in the transformations from algorithm to 
gates can be found. Both these insights into resource estimates and 
evaluation of possible optimizations are enabled with Quipper. 

However, while Quipper paves the way in the direction of a formal framework 
to analyze quantum algorithms, much remains to be done. In particular:
\begin{compactitem}
\item Quipper does not have a dedicated type system. As a result,
 certain programming errors specific to quantum computing, like violating 
 the non-duplicability of quantum registers, remain the responsibility of 
 the programmer.
\item Although one can do concrete gate counts for specific input sizes, it
 is not clear how to automatically get asymptotic gate counts as a
 function of the 
  parameters of the algorithm.
\item Quipper misses a tool to validate that programs do compute
  the correct algorithm.
\end{compactitem}
These questions, together with many others, open a rich and
potentially fruitful research avenue.

\bibliographystyle{eptcs}
\bibliography{generic}
\end{document}